\documentclass[useAMS,usenatbib]{mn2e}
\usepackage{amssymb}
\usepackage{amsmath}
\usepackage{graphicx}
\usepackage{lscape}
\usepackage{color}
\usepackage{subfigure}

\def\h2{\rm{H_2}}
\def\fh2{f_{\rm{H_2}}}

\def\Sh2{\Sigma_{\h2}}
\def\sgas{\Sigma_{\rm{gas}}}

\def\ms{M_{\odot}}
\def\Hs{M_{\h2}/M_*}
\def\HIs{M_{\rm{HI}}/M_*}
\def\HHI{M_{\h2}/M_{\rm{HI}}}
\def\mspc{M_{\odot}\rm{pc^{-2}}}

\title[The Redshift Evolution of Interstellar Metals, Atomic and Molecular Gas]
{The Effect of Star Formation on the Redshift Evolution of Interstellar Metals, Atomic and Molecular Gas in Galaxies }

\author[Fu et al.]{Jian Fu$^{1,2}$ \thanks{E-mail: fujian@mpa-garching.mpg.de; fujian@shao.ac.cn}, Guinevere Kauffmann$^{1}$, Cheng Li$^{2}$, Qi Guo$^{3}$\\
$^1$Max-Planck-Institut f\"ur Astrophysik, D-85741 Garching bei M\"unchen, Germany \\
$^2$Key Laboratory for Research in Galaxies and Cosmology, Shanghai Astronomical Observatory, CAS,\\ 80 Nandan Rd., Shanghai, 200030, China\\
$^3$National Astronomical Observatories, CAS, Beijing 100012, China
}

\begin{document}

\maketitle

\begin{abstract}

We examine how the atomic and molecular gas components of galaxies evolve to higher redshifts using the semi-analytic galaxy formation models of Fu et al. (2010) in which we track the surface density profiles of gas in disks. We adopt two different prescriptions based either on gas surface density and metallicity, or on interstellar pressure, to compute the molecular fraction as a function of radius in each disk. We demonstrate that the adopted star formation law determines how the {\em balance} between gas, stars and metals changes with time in the star-forming galaxy population, but does not influence the total mass in stars formed into galaxies at redshifts below $z\sim 2.5$. The redshift evolution of the mass-metallicity relation places strong constraints on the timescale over which cold gas is converted into stars in high redshift galaxies, and favours models where this remains constant at values around 1-2 Gyr. Future observations of the evolution of the average molecular-to-atomic gas ratio in galaxies as a function of stellar mass and redshift will constrain models of the atomic-to-molecular transition.

\end{abstract}


\section{Introduction}

To understand the formation and evolution of galaxies, we must understand how stars form from the interstellar medium (ISM). The ISM includes gas in ionic, atomic and molecular form. HI in galaxies at low redshift can be probed using the 21cm emission line from neutral hydrogen atoms. HIPASS (HIParkes All Sky Survey; Meyer et al. 2004) and ALFALFA (Arecibo Legacy FAST ALFA Survey; Giovanelli et al. 2005) are the two largest recent ``blind'' 21cm HI surveys covering large areas of sky using single dish radio telescopes. Targeted surveys such as THINGS (The HI Nearby Galaxy Survey, Walter et al. 2008) used the Very Large Array (VLA) to map the atomic gas in a few dozen nearby galaxies, yielding radial surface density profiles for a representative sample of nearby spirals. It is not yet possible to observe 21cm emission from galaxies at redshifts higher than $z\sim0.2$ (Catinella 2008), but damped-Lyman alpha absorbers (DLAs) observed in the
spectra of quasars and gamma ray bursts provide a measure of the integrated atomic hydrogen content at early cosmic epochs (e.g Prochaska \& Wolfe 2009, P{\'e}roux et al. 2003, Savaglio 2006, Noterdaeme et al. 2009). One major drawback of studying DLAs is that the nature of the gas giving rise to the absorption is unclear. It is sometimes postulated that most of the gas is associated with galaxy disks (Kauffmann 1996), but in practice the kinematics are complicated and may indicate that the gas exists in multiple components (e.g Prochaska et al. 2008)

The $\h2$ molecule is very difficult to detect directly, so the CO molecule is generally adopted as a tracer. Single-dish CO surveys of representative samples of nearby galaxies are described in Braine et al. (1993), Young et al. (1995, FCRAO survey), Nishiyama \& Nakai (2001) and most recently, in Saintonge et al. (2011, COLD GASS survey). The BIMA SONG (BIMA survey of Nearby Galaxies, Helfer et al. 2003) and HERACLES (HERA CO-Line Extragalactic Survey, Leroy et al. 2009) mapped the distribution of CO-emitting gas in the same sample of
disk galaxies observed by the THINGS survey (Leroy et al. 2008). In recent years, observations of CO emission from high redshift $z\sim1-3$ strongly star-forming galaxies has become possible (e.g Daddi et al. 2010, Tacconi et al. 2010). There are also observations of CO absorption lines in quasar spectra arising from intervening galaxies at high redshift (e.g Noterdaeme et al. 2009;
Prochaska et al. 2009), but the quasar samples with such measurement are still very small.

Up to a few years ago, most semi-analytic models of galaxy formation (e.g Bower et al. 2006; De Lucia \& Blaizot 2007; Croton et al. 2006; Guo et al. 2011) did not attempt to partition the interstellar cold gas into different phases. Most models adopted a Kennicutt-Schmidt (KS law) star formation law (Kennicutt 1998; Schmidt 1959), in which the star formation rate scaled with the total mass or surface density of cold gas in the galaxy. Galaxy formation modellers have now begun to construct models that include both the atomic and molecular phases of the interstellar gas. The partition of the cold gas into atomic and molecular phases is based either on empirical prescriptions motivated by observations (e.g Blitz \& Rosolowsky 2004 \& 2006), on analytic calculations of the structure of photo-dissociation regions in gas clouds bathed in an external radiation field (Krumholz, Mckee \& Tumlinson 2009a), or on detailed radiative transfer and molecular chemistry calculations embedded within high resolution cosmological simulations of galaxy formation (Gnedin \& Kravtsov 2011). The results derived using all three methods agree with each other at gas densities and metallicities characteristic of the central regions of typical local spirals. They do not extrapolate in the same way to different environments (Fumagalli, Krumholz \& Hunt 2010; Gnedin \& Kravtsov 2011).

Predictions for how molecular gas fractions in galaxies evolve with redshift have been made by taking pre-existing outputs from semi-analytic models and partitioning the cold gas into atomic and molecular components. The density of the stars and the gas in each system is used to calculate the molecular gas fraction (e.g Obreschkow, Heywood, \& Rawlings 2009; Power et al. 2010). If the star formation rate in a galaxy scales with its molecular gas surface density rather than its total gas surface density (Bigiel et al. 2008), this procedure does not treat the conversion from gas to stars self-consistently. More recent models have included the conversion of atomic gas into molecular gas as a module within the semi-analytic model itself (e.g Fu et al. 2010; Cook et al. 2010; Lagos et al. 2011a) and have adopted a star formation models in which the star formation rate is related to the molecular gas content of the galaxy rather than its total gas content. These models have been used to make self-consistent predictions for the evolution of the atomic and molecular gas components in galaxies.

In our previous work (Fu et al. 2010, hereafter Fu10), we included a series of modifications to the models of De Lucia \& Blaizot (2007, hereafter DLB07), which are run on the dark matter halo merger tree outputs of Millennium Simulation (Springel et al. 2005). The surface density profile of infalling gas in a dark matter halo was assumed to be exponential, with scale radius proportional to the virial radius of the halo times its spin parameter. We traced the conversion of atomic gas into molecular gas as a function of radius in each disk by representing each galactic disk by a set of concentric rings. We adopted two prescriptions to partition the cold gas into HI and $\h2$. One prescription was based on the models of Krumholz et al. (2009a) (hereafter $\h2$ prescription 1), in which the molecular gas fraction $f_{\h2}$ is a function of local cold gas surface density and gas metallicity. In $\h2$ prescription 2, $f_{\h2}$ is related to interstellar pressure as described in Elmegreen (1989 \& 1993) and Blitz \& Rosolowsky (2006). Finally, we adopted the star formation law of Leroy et al. (2008) and Bigiel et al. (2008) to relate the star formation rate surface density to the molecular gas surface density.

We showed that the model could fit the HI, $\h2$, and stellar and gas radial surface density profiles of local disk galaxies observed as part of the THINGS/HERACLES surveys, as well as the global stellar, HI and H$_2$ mass functions at $z=0$. A subsequent analysis by Kauffmann et al. (2012) showed that the models also provided a good fit to the relations between atomic and molecular gas mass fraction and stellar mass/surface density in local star-forming galaxies (Saintonge et al. 2011).

In this paper, we will extend this work by examining how the atomic and molecular gas components of galaxies evolve to higher redshifts. In particular, we will study how assumptions about the scaling of the conversion rate from gas to stars in high redshift galaxies affect key observables, including the redshift evolution of the atomic and molecular gas mass fractions of galaxies and the relation between gas-phase metallicity and stellar mass. We will show that the evolution of the mass-metallicity relation places strong constraints on the redshift scaling of the timescale over which gas is converted into stars.

Our paper is organized as follows. In section 2, we will describe our new aspects to the Fu10 models. In section 3, we will present the predictions for the evolution of atomic and molecular gas, star formation, gas-phase metallicity stellar mass in galaxies. We examine the effects of different star formation laws on the model results. We show that the adopted star formation law determines how the {\em balance} between gas, stars and metals changes with time in the galaxy population, but nearly does not influence the total mass in stars formed in galaxies at redshifts below $z=2\sim3$. In section 4, we will summarize the paper and discuss our ideas for future work.

\section{The semi-analytic models}

The semi-analytic models of Fu10 are a modified and extended version of the models in DLB07. The details of the treatment of the gas-physical processes can be found in Sections 2 and 3 in Fu10 and Section 3 in Croton et al. (2006). In this section, we describe the new aspects of the modeling relevant to this paper.

\subsection{Millennium I and II halo merger trees}

The models described in Fu10 were run on the halo outputs of the Millennium Simulation (hereafter MS), which gives reliable results for galaxies with stellar masses $M_*\gtrsim10^{9.5}\ms$. In this paper, we will also use outputs generated from the 125 times higher resolution Millennium II Simulation (Boylan-Kolchin et al. 2009, hereafter MS-II). This is needed to obtain reliable results for lower mass galaxies at high redshifts. In this paper, we run our models on MS-II to get more precise results for galaxies with stellar masses
in the range $10^9\ms<M_*<10^{10}\ms$. We partition the halos in the simulation into two mass ranges: (i) For galaxies in haloes with $M_{\rm halo}>1.72\times10^{10}\ms h^{-1}$ ($1.72\times10^{10}\ms h^{-1}$ is the mass of the minimally resolved halo in the MS), we combine the model results from both MS and MS-II. Since the simulation volume for the MS is 125 times larger than that of MS-II, the model results for these galaxies are actually dominated by MS haloes. (ii) For galaxies in haloes smaller than $1.72\times10^{10}\ms h^{-1}$, we only use results generated using the MS-II haloes.

\subsection{Star formation models} \label{chap:sfrlaw}

The physical processes that regulate the formation of individual stars occur on scales that are very much smaller than that of their host galaxy. Even very high resolution N-body + hydrodynamical simulations of individual galaxies do not resolve the formation of single stars. As a result, all galaxy formation models are required to implement simple prescriptions to model the formation of stars from gas. Ideally, these prescriptions should be motivated both by physical arguments as well as by observations.

In this paper, we adopt a number of different star formation models and we study how they influence the predicted properties of high
redshift galaxies, including their stellar masses, star formation rates, metallicities and atomic and molecular gas fractions. Where possible, we also attempt to make contact with existing observations.

In Fu10, we adopted a model based on the observational studies of Bigiel et al. (2008) and Leroy et al. (2008) to model star formation in galactic disks. These authors found that $\h2$ forms stars at a roughly constant efficiency over the region of the disk where molecular gas dominates. In the outer region of the disk where atomic gas dominates, the conversion efficiency of cold gas into stars was lower and a large variation in $\Sigma_{\rm gas}/\Sigma_{\rm SFR}$ was observed. Using the results presented in these papers, Fu10 adopted the following equations in their models:
\begin{equation}\label{eq:fu10}
\Sigma_{\rm{SFR}} =
\begin{cases}
  \varepsilon\Sigma_{\rm{H_2}} & (f_{\rm{H_2}}\ge 0.5) \\
  \varepsilon'\Sigma_{\rm{gas}}^2 & (f_{\rm{H_2}} < 0.5) \\
\end{cases}
\end{equation}
where $\varepsilon$ is the inverse of molecular gas consumption timescale, and $\varepsilon'=0.5\varepsilon/\Sigma_{\rm gas}|_{\fh2=0.5}$,
so that the star formation rate efficiency changes continuously at the radius where $\fh2=0.5$. The motivation for the change in star formation efficiency at this radius is discussed in detail in Fu10. In the radially resolved multiple ring model, the gas surface density and stellar surface density are azimuthally averaged in each ring. In the outer parts of galaxy disks, the mean surface density and mean molecular fraction are very low. In these regions in real galaxies, the star formation tends to be dominated by small isolated regions where the local surface density is much higher than the azimuthally averaged value, and as a result the local molecular
fraction is higher. Since we cannot easily capture this effect in our azimuthally averaged prescription, we instead modify the star formation law based on an empirical fit to the $\Sigma_{\rm SFR}$ vs $\Sigma_{\rm gas}$ relation in low gas surface density regions in Bigiel et al. (2008). If we do not do this, our models predict gas surface densities in the outer disk that are too high (see Sec. 3.4 par. 1 to 3 in Fu10).

Note that the prescription in Eq. (\ref{eq:fu10}) is based on an average fit to the data, and does not account for the observed {\em scatter} between different regions of the disk, or between different galaxies (Saintonge et al. 2011). In this paper, we are concerned with the evolution of the average properties of galaxy populations, so we will neglect these complications.

The star formation models in both DLB07 and Croton et al. (2006) are based on the treatment of star formation in galactic disks outlined in Kauffmann (1996), which was motivated by the 1989 study of star formation by Kennicutt. Kennicutt found that the radial profiles of star formation in nearby spiral galaxies could be explained if the star formation rate surface density was proportional to a power $N$ (with $N$=1-2) of the total gas surface density in the inner regions of disks (i.e. a ``Schmidt-law''), followed by a cut-off in the outer regions of the disk. The cut-off was explained by simple gravitational stability considerations based on the Toomre (1964) calculation of
the stability of thin isothermal gas disks.

The Kauffmann (1996) semi-analytic model assumed that all disks have constant gas velocity dispersion of 6 km/s and flat rotation curves
with rotational velocity equal to the circular velocity of the surrounding dark matter halo. In this case, the critical surface density for star formation can be written
\begin{equation}\label{eq:sigmacrit}
\Sigma_{\rm{crit}}=0.59{M_\odot}{\rm{pc}}^{-2}\left(\frac{v_{\rm{vir}}}{\rm{km~s}^{-1}}\right)\left(\frac{r}{\rm{kpc}}\right)^{-1}
\end{equation}

For gas at surface densities greater than $\Sigma_{\rm crit}$, Kauffmann (1996) did not adopt a Schmidt law, but a relation of the
form $\Sigma_{\rm SFR}\propto\Sigma_{\rm gas}/t_{\rm dyn}$, where $t_{\rm dyn}$ is the dynamical timescale of the disk, given by $R_{\rm disk}/V_{\rm c}$. Global star formation laws where the star formation surface density scales with the ratio of the gas density to the dynamical time were first proposed by Wyse (1986) and Wyse \& Silk (1989). They are expected if star formation is regulated by dynamical structures such as bars or spiral arms, leading to a scaling of star formation with the orbital frequency of the galaxy. Kennicutt (1998) tested both the standard Schmidt law as well as a star formation law with an orbital time dependence. He concluded that both provided equally good fits to a sample of normal spirals and IR-selected star burst galaxies. More recently, Bigiel et al. (2008) tested whether the conversion efficiency of the {\em molecular gas} into stars (as opposed to the total cold gas) depended on the orbital frequency and found no effect.

In this paper, we investigate two star formation models in which the star formation rate scales with the orbital time in the galaxy. In the first such model, the star formation rate scales with the total surface density of cold gas that is above the critical threshold for star formation divided by the orbital timescale. We adopt
\begin{equation}\label{eq:dlb07}
\Sigma_{\rm{SFR}} =
\begin{cases}
 0 & \left( \sgas < \Sigma_{\rm crit} \right) \\
 \alpha\left(\sgas-\Sigma_{\rm crit}\right)/t_{\rm dyn} & \left(\sgas \ge \Sigma_{\rm crit} \right) \\
\end{cases}
\end{equation}
in which $t_{\rm dyn}$ and $\Sigma_{\rm crit}$ are calculated as indicated above.

In the second such model, the star formation rate scales with the molecular gas surface density divided by the orbital timescale. A recent formulation of this was given in Genzel et al. (2010), where the relation between galaxy-averaged star formation rate and global molecular gas mass surface density was studied for star-forming galaxies at $z\sim1-3$. The relation between molecular gas mass surface density and star formation rate was well fit by the functional form
\begin{multline}\label{eq:genzelsfr2}
\log \left[\frac{\Sigma_{\rm SFR}}{\ms \rm{yr^{-1} kpc^{-2}}} \right] = \\
-1.76(\pm0.18)+0.98(\pm0.09)\log \left[ \frac{\Sigma_{\rm{molgas}}/\tau_{\rm{dyn}}}{\ms \rm{yr^{-1} kpc^{-2}}} \right]
\end{multline}
in which $\Sigma_{\rm SFR}$, $\Sigma_{\rm mol}$ are disk-averaged values and $\tau_{\rm dyn}$ is the dynamical timescale of the disk.
We adopt a slightly simplified version of the form
\begin{equation}\label{eq:genzel10}
\Sigma_{\rm SFR} =\alpha \Sh2/t_{\rm dyn}
\end{equation}
where $\alpha$ in Eq. (\ref{eq:genzel10}) is a tunable parameter and
where $t_{\rm dyn}$ is the global dynamical time of the disk. $\Sigma_{\rm SFR}$ and $\Sh2$ is the local surface density within one of the radial rings of the disk. We note that linear superposition of the relation applied to each ring does lead to recovery of the global relation. There is, however, no current empirical evidence that the relation hold on scales smaller than that of the entire galaxy, so we caution that our attempt to apply this relation to our radial ring prescription is something of an extrapolation.

Krumholz, Mckee \& Tumlinson (2009b) proposed a theoretical model for the local star formation rate in a galaxy based on the following three assumptions: a) the fraction of gas that is eligible to form stars is the fraction that is in molecular form; b) the properties of molecular clouds and hence the rate they form stars is independent of galaxy properties until the galactic ISM pressure becomes comparable to the internal pressure within molecular clouds; c) above this limit, molecular cloud properties and the rate at which they form stars scales with ISM pressure. They then derive a star formation ``law'' of the form:
\begin{equation}\label{eq:kmt09}
\Sigma_{\rm SFR}=
\begin{cases}
  \varepsilon\Sh2\left(\Sigma_{\rm{gas}}/\Sigma_0\right)^{-0.33}& \left(\Sigma_{\rm{gas}}<{\Sigma_0}\right)\\
  \varepsilon\Sh2\left(\Sigma_{\rm{gas}}/\Sigma_0\right)^{ 0.33}& \left(\Sigma_{\rm{gas}}>{\Sigma_0}\right)\\
\end{cases}
\end{equation}
in which $\varepsilon=0.39~\rm{Gyr^{-1}}$, and $\Sigma_0=85\mspc$. The division into low and high surface density regimes at $85\mspc$ is adopted so that the model is able to fit the observed $\dot{\Sigma_*}$ versus $\Sigma_{\rm gas}$ relation for both ``normal'' spiral galaxies and for samples of nearby star burst galaxies, for which $\Sigma_{\rm gas}$ values are higher and the observed conversion time of molecular gas into stars is shorter. We note that at high redshifts, typical disk galaxies are more compact and have higher surface densities of gas, i.e. they are more similar to present-day starbursts.

We will explore whether the four star formation prescriptions outlined above can be differentiated via the predicted redshift evolution of the relations between gas, stellar mass and gas-phase metallicity in galaxies. In the standard picture of disk galaxy formation in a
hierarchical Universe, the sizes of galaxies of fixed circular velocity decrease at higher redshifts, so the disk dynamical times are shorter (Mo, Mao \& White 1998). Because gas cooling rates are higher at early epochs, galaxies are more gas rich. However, because galaxies are also smaller at high redshifts, gas surface densities are higher and the interstellar medium will contain a higher fraction of gas in the molecular phase, so star formation can be more efficient. Finally, if star formation rates scale with the surface density of molecular gas divided by the dynamical time, gas consumption times will be shorter in high redshift galaxies than locally, and the rate of
metal enrichment of the ISM will be higher. We would thus expect the redshift evolution of the relations between star formation, stellar mass, gas-phase metallicity and gas fraction to be quite sensitive to the adopted star formation model.

From now on, we will refer to the star formation prescription in Eq. (\ref{eq:fu10}) as ``Bigiel'', to that in Eq. (\ref{eq:dlb07}) as ``Kennicutt'', to that in Eq. (\ref{eq:genzel10}) as ``Genzel'' and to that in Eq. (\ref{eq:kmt09}) as ``Krumholz''.

\subsubsection{Tuning the model parameters } \label{chap:h2mfcoldgass}

The three star formation laws that depend on molecular gas surface densities (Bigiel, Genzel and Krumholz) are run with the two $\h2$ fraction prescriptions described in Fu10. Following the nomenclature adopted in that paper, the prescription based on the analytic models of Krumholz et al. (2009a) will be referred to as ``prescription 1'', while the prescription based on interstellar pressure will be referred to as ``prescription 2''. As we will show, both prescriptions yield nearly identical results at $z=0$ if the star formation model is held fixed.

Our procedure is to first adjust the free parameters of the model to fit the observed stellar, HI, $\h2$ and cold gas mass functions at $z=0$. In the case of the Kennicutt star formation model, we only fit the stellar and the cold gas mass function. Table 1 in Fu10 lists the free parameters of the models that we adjusted to fit this data. We note that Fu10 adopted the $\h2$ mass function derived by Keres et al. (2003) from the FCRAO Extragalactic CO survey in which a CO-$\h2$ conversion factor of $X=3\times10^{20}\rm{cm^{-2}~K^{-1}~km^{-1}~s}$ as adopted. Note that $X$ is defined as
\begin{equation}\label{eq:xco}
X=\frac{{{N_{\h2}}/{\rm{cm}}^{-2}}}{{{I_{{\rm{co}}}}/{\rm{K~km~s}}^{-1}}}
\end{equation}
More recent surveys have adopted $X=2\times10^{20}\rm{cm^{-2}~K^{-1}~km^{-1}~s}$ (or $\alpha_{\rm CO}=3.2\ms/\rm{K~km~s^{-1}~pc^2}$, Leroy et al. 2008; Bigiel et al. 2008; Saintonge et al. 2011). In order to make our models consistent with this data, it is sufficient to increase the supernovae heating rate by a relatively small factor (for example $\epsilon_{\rm disk}$ increases from 3.5 to 5.0 for the Bigiel model). An increase in $\epsilon_{\rm disk}$ leads to a downwards shift the total cold gas-to-stellar mass fraction and compensates for the change in $X$.

\begin{figure*}
\centering
 \includegraphics[angle=-90,scale=0.60]{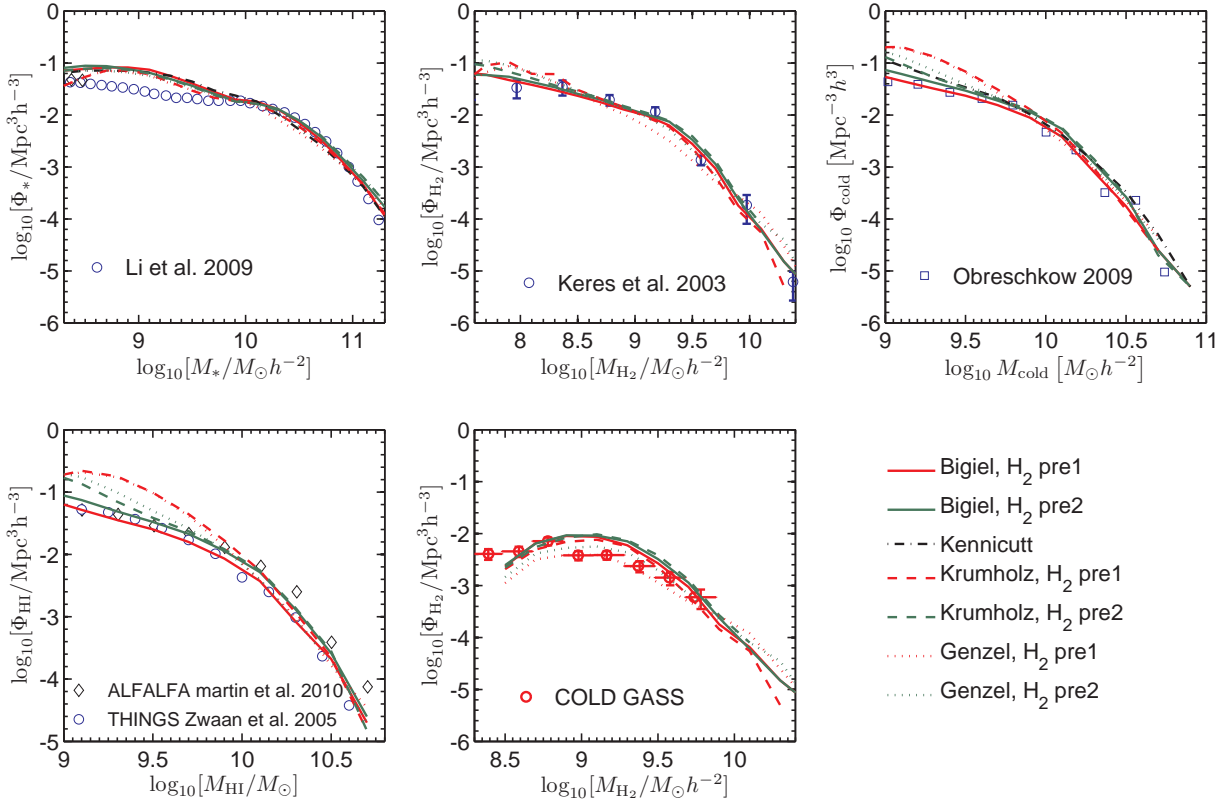}\\
 \caption{HI, $\h2$, total cold gas mass and stellar mass functions at $z=0$ in the models compared with the observations. The observed stellar mass function is from Li et al. (2009). Open circles show the HI mass function from Zwaan et al. (2005) and open diamonds show the HI mass function from Martin et al. (2010). The $\h2$ mass function data is from Keres et al. (2003) and from COLD GASS (see text).
 The top right panel shows the total cold gas mass function derived by Obreschkow \& Rawlings (2009). In each panel, different curves are for the star formation models and $\h2$ fraction prescriptions as indicated in the right side of this figure.
} \label{fig:mfz0}
\end{figure*}

In Fig. \ref{fig:mfz0}, we present the $z=0$ HI, $\h2$, cold gas and stellar mass functions for all our models and compare them with
observations. The observed stellar mass function shown in the top left panel is derived from the data release 7 of the Sloan Digital Sky Survey (SDSS) (Li et al. 2009). The HI mass functions shown in the bottom left panel are derived from the HI Parkes All-Sky Survey (HIPASS) survey (Zwaan et al. 2005; open diamonds) and the Arecibo Legacy Fast ALFA Survey (ALFALFA) (Martin et al. 2010; open circles). The $\h2$ mass function in the top middle panel is from the Five College Radio Astronomy Observatory (FCRAO) survey (Keres et al. 2003). The total cold gas mass function in the top right panel is from the combination of the HIPASS HI and FCRAO $\h2$ mass functions by Obreschkow \& Rawlings (2009). Note that $X=2\times10^{20}\rm{cm^{-2}~K^{-1}~km^{-1}~s}$ has been adopted uniformly throughout.

In the bottom right panel, we plot the $\h2$ mass functions derived for galaxies with $M_*> 10^{10}\ms$ with $\h2$ detections in the COLD GASS survey (Saintonge al. 2011). In order to correct for the mass-dependent selection of the COLD GASS, as well as to estimate an effective volume of the survey, we have selected a volume-limited sample of 24,000 galaxies with $M_*>10^{10}\ms$ and $0.025<z<0.05$ from the SDSS data release 7. Since we have full understanding of the survey geometry and masks, it is easy to compute the effective volume occupied by this sample. COLD GASS targets a random subset of galaxies with stellar masses greater than $M_*>10^{10}\ms$. The COLD GASS galaxies with CO line detections are placed in bins of stellar mass of width 0.2 dex and assigned a weight, given by the ratio between the total number of galaxies in SDSS sample and the total number of COLD GASS galaxies in the same mass bin. The number density of galaxies for a given $\h2$ mass range is then estimated by the total weight of the galaxies in that bin divided by the volume of the SDSS sample. The $\h2$ mass function determined in this way is listed in Tab. (\ref{tab:h2mfcoldgass}) and the $\h2$ mass function from galaxies with CO line detections (column 4 \& 5) are plotted in the bottom right panel of Fig. \ref{fig:mfz0}. The errors indicate the Poisson noise in the sample with detections.

\begin{table*}
 \centering
 \caption{The $\h2$ mass function derived from COLD GASS sample (Saintonge et al. 2011). Column 1, 2 are the lower, upper values of $\h2$ mass bins. Column 4 \& 5 are the $\h2$ mass functions and errors if we only include galaxies with CO line detections; column 7 \& 8 are the $\h2$ mass function and errors if we also include galaxies without CO line detections and assign them an $\h2$ mass equal to the upper limit. Column 3 \& 6 are the mean $\h2$ mass of the galaxies sample with and without CO line detections. $h=0.7$ is assumed for both mass and distance calculations, and the adopted CO-$\h2$ conversion factor is $X=2\times10^{20}\rm{cm^{-2}~K^{-1}~km^{-1}~s}$.
 }\label{tab:h2mfcoldgass}
 \begin{tabular}{|c|c|c|c|c|c|c|c|}
 \hline \hline
  $\log_{10}M_{\h2}^{\rm lower}$ &  $\log_{10}M_{\h2}^{\rm upper}$ & $\log_{10}M_{\h2}^{\rm mean}$ & $\Phi$ &  error  & $\log_{10}M_{\h2}^{\rm mean}$ &  $\Phi$    & error \\
  ($\ms$) &  ($\ms$)    &($\ms$) &($\rm{Mpc^{-3}~dex^{-1}}$)&($\rm{Mpc^{-3}~dex^{-1}}$)&($\ms$)&($\rm{Mpc^{-3}~dex^{-1}}$)&($\rm{Mpc^{-3}~dex^{-1}}$)\\
 \hline
8.35&8.55 & 8.49& 5.05e-4 &2.26e-4& 8.48& 1.63e-3 &4.07e-4 \\
8.55&8.75 & 8.66& 1.59e-3 &3.74e-4& 8.65& 5.14e-3 &6.81e-4 \\
8.75&8.95 & 8.86& 1.77e-3 &3.97e-4& 8.85& 3.66e-3 &5.29e-4 \\
8.95&9.15 & 9.05& 2.79e-3 &4.78e-4& 9.04& 4.47e-3 &5.03e-4 \\
9.15&9.35 & 9.25& 1.49e-3 &3.04e-4& 9.25& 1.68e-3 &2.97e-4 \\
9.35&9.55 & 9.43& 1.51e-3 &2.85e-4& 9.43& 1.57e-3 &2.92e-4 \\
9.55&9.75 & 9.64& 9.23e-4 &2.24e-4& 9.64& 9.24e-4 &2.24e-4 \\
9.75&9.95 & 9.84& 5.55e-4 &1.60e-4& 9.84& 5.55e-4 &1.60e-4 \\
9.95&10.15&10.01& 2.32e-4 &9.48e-5&10.01& 2.32e-4 &9.48e-5 \\
\hline \hline
\end{tabular}
\end{table*}

The red and green curves in Fig. \ref{fig:mfz0} show results for the four star formation models described in the previous section. For the models that involve molecular gas (Bigiel, Genzel and Krumholz), we how results for both $\h2$ fraction prescriptions. For the Kennicutt model, we only compare the {\em total} cold gas mass function with observations. Note that when we compare the models with the $\h2$ mass function derived from the COLD GASS survey, we select model galaxies that fall within the COLD GASS detection limits: i.e. galaxies are included if $\log_{10}\left[M_{\h2}/M_*\right]\le-1.72$ if $M_*>10^{10.6}\ms$, and $\log_{10}\left[M_{\h2}/M_*\right]\le8.78-\log_{10}[M_*/\ms]$ if $10^{10} < M_*/\ms<10^{10.6}$ (Kauffmann et al. 2012).

Fig. \ref{fig:mfz0} shows that the $\h2$ and stellar mass functions for all the models are nearly identical. Recall that we tune our model parameters to match the normalization of these functions, so this is not entirely surprising. The low mass end of the stellar mass function is steeper than in the observations. A better fit requires changes to the supernova feedback model itself (see for example Guo et al. 2011 for a recent discussion). We will come back to this matter in the final section of the paper. All the models provide a good fit to the FCRAO mass function. The bottom right panel shows that the models also yield a reasonable fit to the $\h2$ mass function derived from COLD GASS.

At low redshifts, the largest differences between models arise at the low mass end of the HI mass function. The HI and cold gas mass functions obtained using the KMT and Genzel star formation models, appear to disagree with the observed HI mass function at the low mass end, particularly if $\h2$ prescription 1 is adopted. In contrast, the HI mass function obtained using the Bigiel star formation model Eq. (\ref{eq:fu10}) is in somewhat better agreement with observations. As seen from Eq. (\ref{eq:fu10}), stars will form even when the cold gas {\em in a given radial ring} is dominated by atomic hydrogen. This is not the case in our adaptation of the Genzel and the Krumholz models. In low mass galaxies, the gas metallicity is low (see Sec. \ref{chap:mzrelation}). If we adopt $\h2$ prescription 1, where the threshold density for gas to be converted from atomic to molecular form depends on metallicity, a larger fraction of the gas will not be able to form stars in the Genzel and Krumholz models, leading to an excess at the low mass end of the HI mass function. Similar results were found by Lagos et al. (2011a, Figure 6). We caution that our model does not include detailed treatment of photon-ionization processes. In low mass galaxies, such processes are more important (e.g Gnedin \& Kravtsov 2011) and part of the atomic gas will actually be in the form of ionized HII. This may bring the predicted HI mass function into better agreement with observations.

\begin{table*}
 \centering
 \caption{The model parameters for the four star formation models. $\alpha$ is the star formation efficiency in Eq. (\ref{eq:dlb07}) \& (\ref{eq:genzel10}).
 $\varepsilon$ is the inverse of molecular star formation time scale in Eq. (\ref{eq:fu10}) \& (\ref{eq:kmt09}).
 $\kappa_{\rm BH}$ is the quiescent hot gas black hole accretion rate. $\epsilon_{\rm disk}$ is the supernova reheating rate.
 }\label{tab:parameter}
 \begin{tabular}{|c|c|c|c|c|}
 \hline \hline
star formation law & $\alpha$ & $\varepsilon[10^{-10}\rm{yr}^{-1}]$ & $\kappa_{\rm BH}[10^{-6}\ms\rm{yr^{-1}}]$ & $\epsilon_{\rm disk}$ \\
 \hline
Bigiel & &$4.9 $ & 2.0 & 5.0 \\
Krumholz & & $3.9$ & 2.0 & 5.0 \\
Kennicutt & 0.021 & & 5.0 & 5.0 \\
Genzel & 0.032 & & 5.0 & 5.0 \\
\hline \hline
\end{tabular}
\end{table*}

In Tab. (\ref{tab:parameter}), we list the parameter values adopted for our four star formation models. The values for the Bigiel star formation are identical to those used in Fu10 except for the rescaling of $\epsilon_{\rm disk}$ described above. In the other three cases, we adjust a combination of the star formation efficiency parameter $\alpha$ and the quiescent hot gas black hole accretion rate $\kappa_{\rm BH}$ to bring the $z=0$ stellar and gas mass functions into agreement with observations. We note that the molecular gas consumption time $\varepsilon$ is fixed to the values given in the Bigiel et al. (2008) and Krumholz et al. (2009b) papers. Other model parameters are the same as in Table 1 of Fu10.

\section{Model results at high redshift}

In this section, we will compare results at high redshifts for the four different star formation models described above. In Sec. \ref{chap:sfrevolution}, we will discuss the redshift evolution of the cosmic star formation rate density and compare results with recent observations. In Sec. \ref{chap:mzrelation}, we will investigate the redshift evolution of the relation between gas-phase metallicity and stellar mass (MZ relation). In Sec. \ref{chap:gasratios}, we will study the evolution of atomic and molecular gas fractions and molecular-to-atomic gas ratios in galaxies, and investigate predictions for the evolution of the CO luminosity function. In Sec. \ref{chap:smfz}, we will discuss the redshift evolution of stellar mass function.

\subsection{The evolution of cosmic star formation rate density} \label{chap:sfrevolution}

\begin{figure*}
\centering
 \includegraphics[angle=-90,scale=0.75]{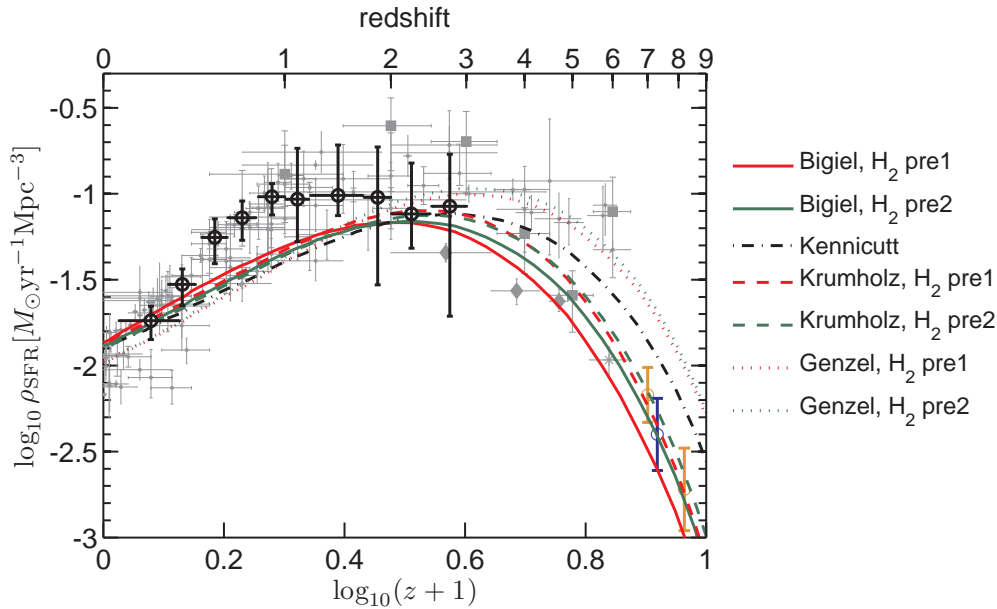}\\
 \caption{The evolution of the cosmic star formation rate for the models is compared with the observations. The grey data points with error bars are from the compilation of Hopkins et al. (2007). Orange data points are from Bouwens et al. (2010); blue data points are from Bouwens et al. (2008). The curves show model results (solid curves are for the Bigiel model; dashed curves are for the Krumholz model; dotted curves are for the Genzel model; dash-dot curves are for the Kennicutt model). Red curves and green curves indicate $\h2$ prescription 1 and 2 respectively. For the Kennicutt model, results have been plotted in black.}
\label{fig:sfrz}
\end{figure*}

The evolution of cosmic star formation rate density (commonly known as the ``Madau plot'', Madau, Pozzetti \& Dickinson 1998) is frequently used to constrain galaxy formation models. Fig. \ref{fig:sfrz} compares our model results with the data. The gray data points with error bars are from the compilation of Hopkins et al. (2007), while the orange and blue data points at $z>7$ are derived from the z-dropout and Y-dropout galaxies analyzed by Bouwens et al. (2008 \& 2010). The curves show model results. For the three star formation laws involving molecular gas, we show results for our two $\h2$ prescriptions.

From Fig. \ref{fig:sfrz}, we can see that the model results are all very similar out to redshifts of $\sim 2.5$ and then diverge. Because galaxies are more compact at high redshift, their global dynamical timescales are shorter. According to DLB07,
\begin{equation}\label{eq:tdynz}
t_{\rm dyn}\propto r_{\rm d}/v_{\rm vir}\propto\lambda r_{\rm vir}/v_{\rm{vir}}\propto\left(1+z\right)^{-3/2}
\end{equation}
where $r_{\rm d}$ is the disk scale length, $v_{\rm vir}$ and $r_{\rm vir}$ are the virial velocity and virial radius of the surrounding halo, and $\lambda$ is the spin parameter of of the halo. The Kennicutt and Genzel models have a $1/t_{\rm dyn}$ scaling, so that at fixed gas surface density, the star formation rate will be proportional to $(1+z)^{3/2}$ in DLB07 and Fu10 models. The star formation rate densities predicted by models in which SFR scales with $1/t_{\rm dyn}$ (i.e. Kennicutt and Genzel models) are thus higher, particularly at $z> 3$. In the Krumholz model, the conversion rate of gas into stars is higher at large gas surface densities. Because high-redshift galaxies are more gas-rich (see Sec. \ref{chap:h2fracz}), the Krumholz model yields slightly higher SFR densities at high $z$ than the Bigiel model. We note, however, that the effect is not a strong one, because the scaling of the molecular gas conversion efficiency with gas surface density is relatively weak.

We also see that the high redshift star formation rate densities with $\h2$ prescription 2 are all a bit higher than with $\h2$ prescription 1. As will be shown in Section \ref{chap:h2fracz}, this is caused by the fact that $\h2$ prescription 1 leads to lower molecular-to-atomic gas fraction ratios at high redshift.

When we compare the model results to the observations, we see that none of the models reproduce the very steep apparent rise in cosmic SFR density from $z=0$ to $z=2$. This problem has already been remarked upon in a number of papers in the literature (e.g. Guo et al. 2011, hereafter Guo11). Recently, there have been analyses based on Herschel far-IR data showing that star formation rates based on 24 $\mu$m fluxes may be systematically overestimated at $z>0.7$ by factors that are similar to the discrepancy shown in the figure (Nordon et al. 2012). A preliminary recalibration of the Madau plot is given in Gruppioni et al. (2010), based on deep observations of the GOODS-N field.
Results from this paper are plotted as black data points in Fig. \ref{fig:sfrz}. As can be seen, the discrepancy with the models is no longer present at $z \sim 2$, but persists at lower redshifts. We note, however, that the GOODS-N is quite small and Gruppioni were required to extrapolate their IR-luminosity functions by large factors to obtain their results.

At redshifts greater than 6, the Bigiel and Krumholz models yield results that are in better agreement with currently available data.

\subsection{The relations between gas-phase metallicity and stellar mass as a function of redshift} \label{chap:mzrelation}

The gas-phase metallicities of high redshift galaxies are relatively easy to estimate from the strengths of emission lines such as [OII], [OIII] and H$\beta$ observed in their spectra. The relation between stellar mass and metallicity for a large sample of emission line galaxies from the Sloan Digital Sky survey was first quantified by Tremonti et al. (2004), and the evolution of this relation to high redshift has been the subject of many papers. Maiolino et al. (2008) provide parameterized fits to the evolution of the mass-metallicity relation out to $z \sim 3.5$, which we use as the observational basis for the model comparisons in this section.

\begin{figure*}
\centering
 \includegraphics[angle=-90,scale=0.69]{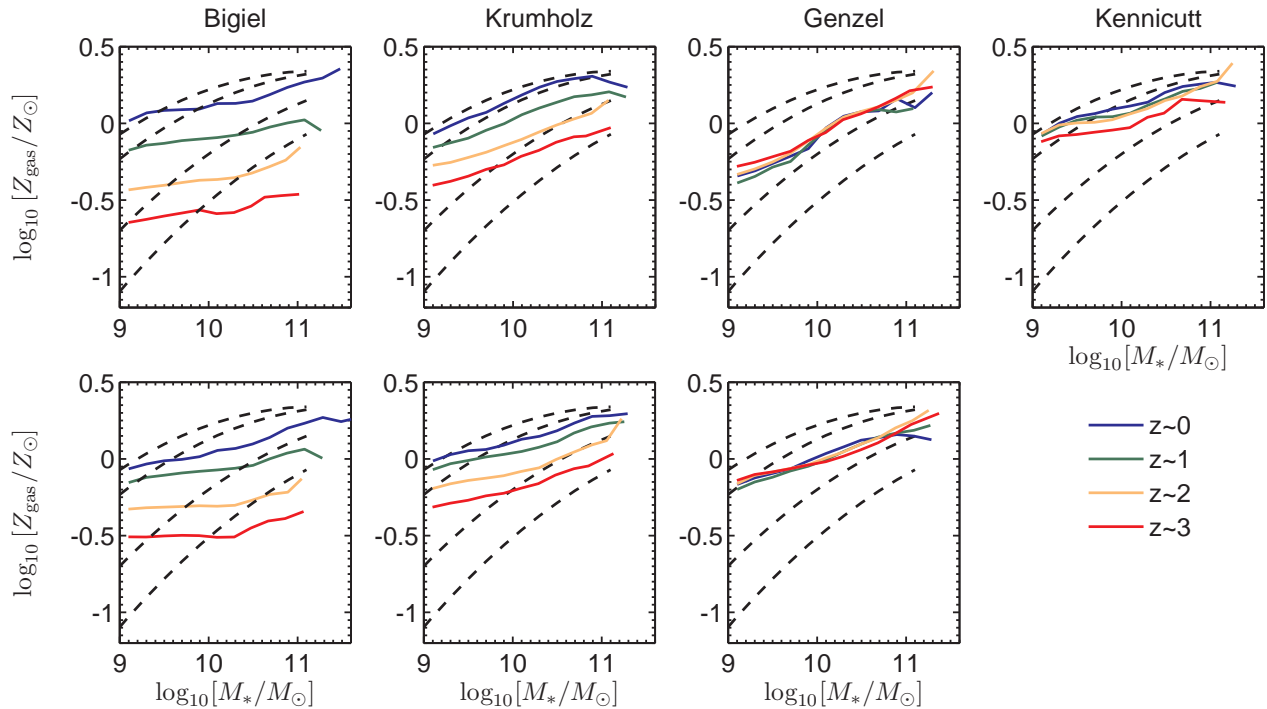}\\
 \caption{The relation between gas-phase metallicity and stellar mass for star-forming galaxies is plotted at different redshifts. The 4
 columns show results for the 4 star formation models described in Section \ref{chap:gasratios}. The two rows show results for $\h2$ prescriptions 1 and 2 (except the Kennicutt star formation model). In each panel, the solid coloured curves show the mean gas-phase metallicity as a function of stellar mass at redshifts $z\sim0,~1,~2,~3$. The dashed curves are the fits of the observational data by Maiolino et al. (2008).
 } \label{fig:mzrelation}
\end{figure*}

In Fig. \ref{fig:mzrelation}, we plot the mean values of gas-phase metallicity as a function of stellar mass and redshift for star-forming galaxies in the models. The definition of ``star-forming'' adopted here is the galaxy $\dot M_*/M_*>10^{-11}\rm{yr^{-1}}$. The four columns in Fig. \ref{fig:mzrelation} show results for the four star formation models; the top and bottom panels show results for $\h2$ fraction prescription 1 \& 2 respectively (for Bigiel, Krumholz, Genzel star formation models). In each panel, the four coloured curves represent model results at $z\sim0,~1,~2,~3$. The dashed curves are the fitting curves compiled by Maiolino et al. (2008) from observations of star-forming galaxies at $z=0.07$ from Kewley \& Ellison (2008), at $z \sim 0.7$ from Savaglio et al. (2005), and at $z\sim2$ from Erb et al. (2006). The redshift 3.5 curve is a fit to the VLT AMAZE survey results of Maiolino et al. (2008). To compare the observed oxygen abundance results to the gas-phase metallicity in our models, we adopt the solar oxygen abundance 12+$\log_{10}\rm{(O/H)}_{\odot}=8.69$ in Asplund et al. (2009) to convert the values of $12+\log_{10}(\rm{O/H})_{\rm gas}$ in Maiolino et al. (2008) to gas phase metallicity in units of the solar value $\log_{10}[Z_{\rm gas}/Z_{\odot}]$ in Fig. \ref{fig:mzrelation}.

We begin by noting that the slope of the mass-metallicity relations predicted by all the models is too shallow compared to observations. This is again a ``problem'' inherited from the DLB07 model. It has been shown by Guo11 that the change in the supernova feedback model
that brings the stellar mass function in better agreement with observations, also steepens the mass-metallicity relation, resulting in a better fit to the observations.

In this paper, we leave the supernova feedback model fixed and focus only on the redshift evolution of the mass-metallicity relation.
As can be seen, the Kennicutt and Genzel star formation models in which the gas consumption time scales with the inverse dynamical time,
leads to a mass-metallicity relation that does not evolve with redshift. In contrast, the Bigiel and Krumholz star formation models
in which the star formation rate depends only on the surface density of molecular gas, produces a much more strongly evolving mass-metallicity relation.

High redshift galaxies have lower gas-phase metallicities than low redshift galaxies of the same mass, because they accrete low metallicity gas from the external medium at higher rates. As we will show in the next section, the ratio of gas-to-stars decreases strongly from high redshift to the present day. In models where the star formation rate scales as $1/t_{\rm dyn}$, the accreted gas consumed into stars much more quickly. As a result, gas-phase metallicities and gas mass fractions evolve less strongly with redshift at $z\lesssim3$.

If we compare our models with the observations, we see that only the Bigiel star formation model yields mass-metallicity relations that evolve as strongly as observed. We caution, however, that more careful comparison is warranted. We have adopted a fixed specific star formation rate threshold of $\dot M_*/M_*>10^{-11}\rm{yr^{-1}}$ in the models to define our star-forming samples at all redshifts. In practice, emission-line detection thresholds may be higher for faint galaxies at higher redshifts. This might lead to the observed samples that are biased towards more gas-rich galaxies at higher redshifts and would result in stronger {\em apparent} evolution of the mass-metallicity relation.

In summary, star formation models with fixed dependence on gas surface density yield high redshift mass-metallicity relations that are in much better agreement with observations than the models where the gas conversion time scales with the orbital timescale of the galaxy.

\subsection{The gas content of high redshift galaxies} \label{chap:gasratios}

In this section, we will present model predictions for the evolution of atomic and molecular gas in galaxies to high redshifts. First,
we will discuss the redshift evolution of the cosmic molecular-to-atomic gas density ratio. Then, we will show model predictions for how the molecular and atomic gas mass fractions ($m_{\h2}/m_*$ and $m_{\rm HI}/m_*$) in galaxies of different stellar masses evolve to higher redshifts. Finally, we will present predictions for the CO luminosity function at different redshifts, first assuming that the conversion between $\h2$ mass and CO luminosity is the same in all galaxies, and then adopting the metallicity-dependent conversion proposed by Feldmann, Gnedin \& Kravtsov (2012) and Narayanan et al. (2011).

\subsubsection{The evolution of gas mass-to-stellar mass and molecular-to-atomic gas ratios} \label{chap:h2fracz}

\begin{figure*}
\centering
\includegraphics[angle=0,scale=0.5]{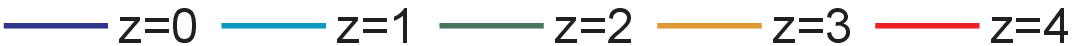}\\
\includegraphics[angle=0,scale=0.40]{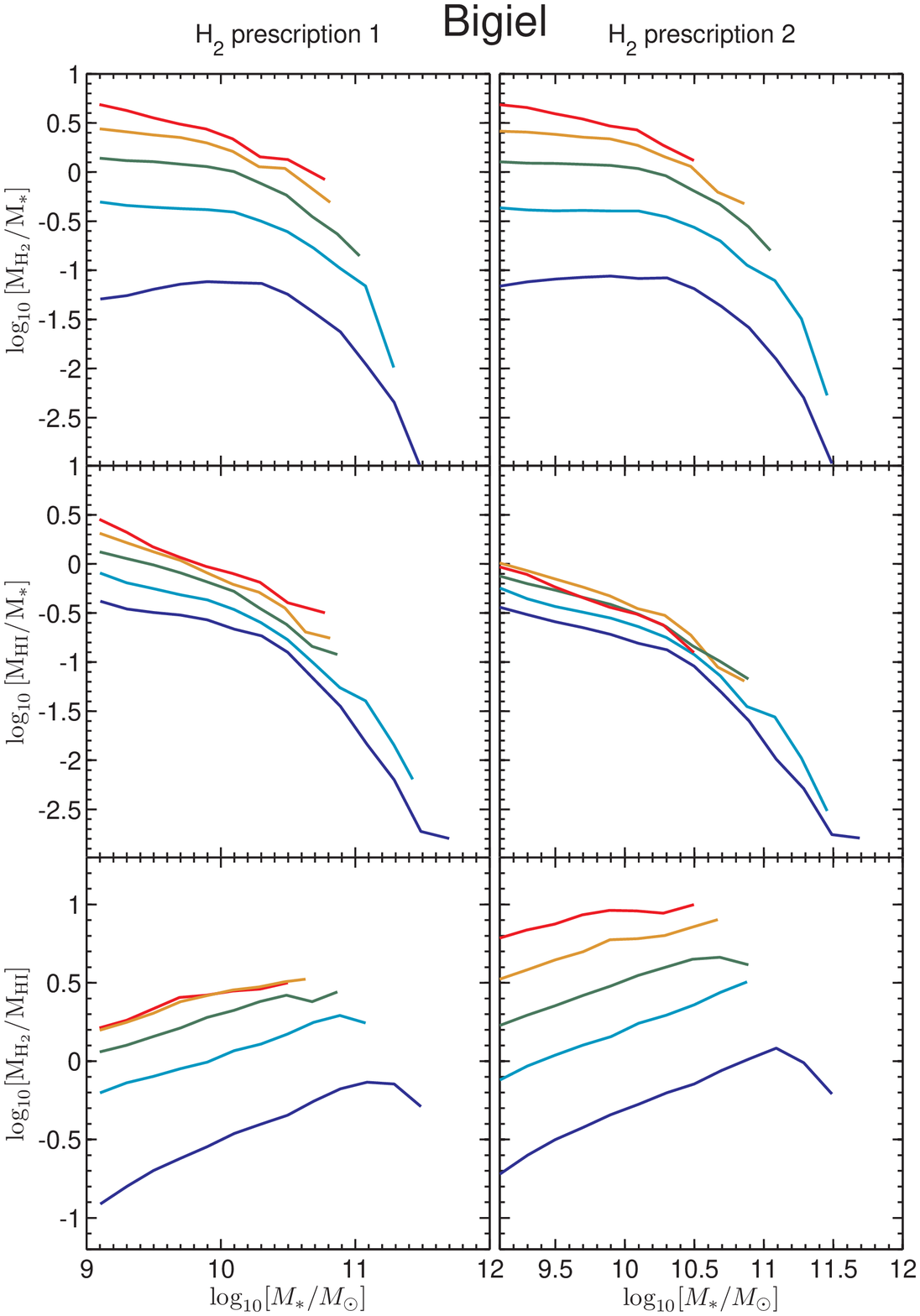}
\includegraphics[angle=0,scale=0.40]{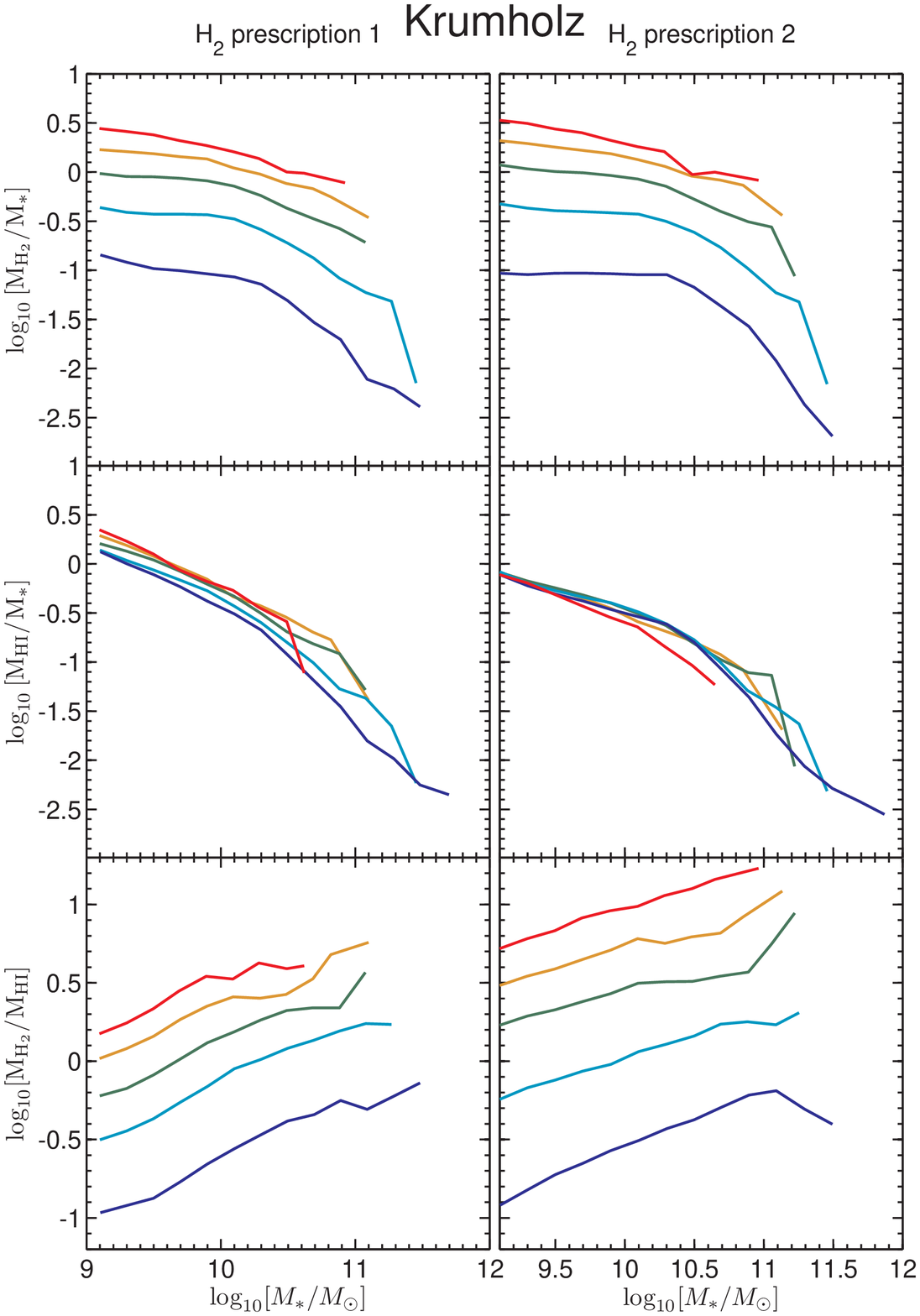}\\
\includegraphics[angle=0,scale=0.40]{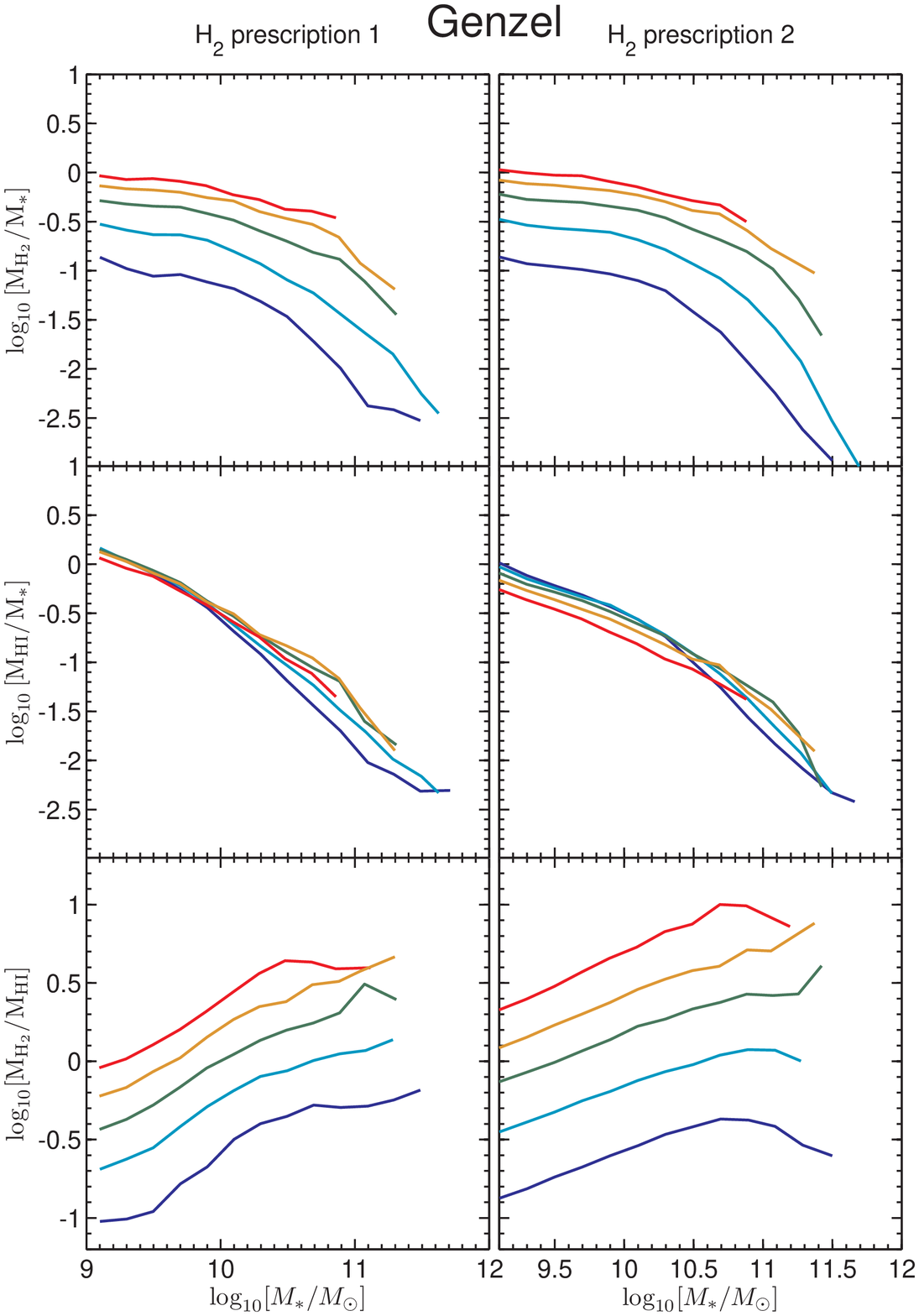}
\includegraphics[angle=0,scale=0.42]{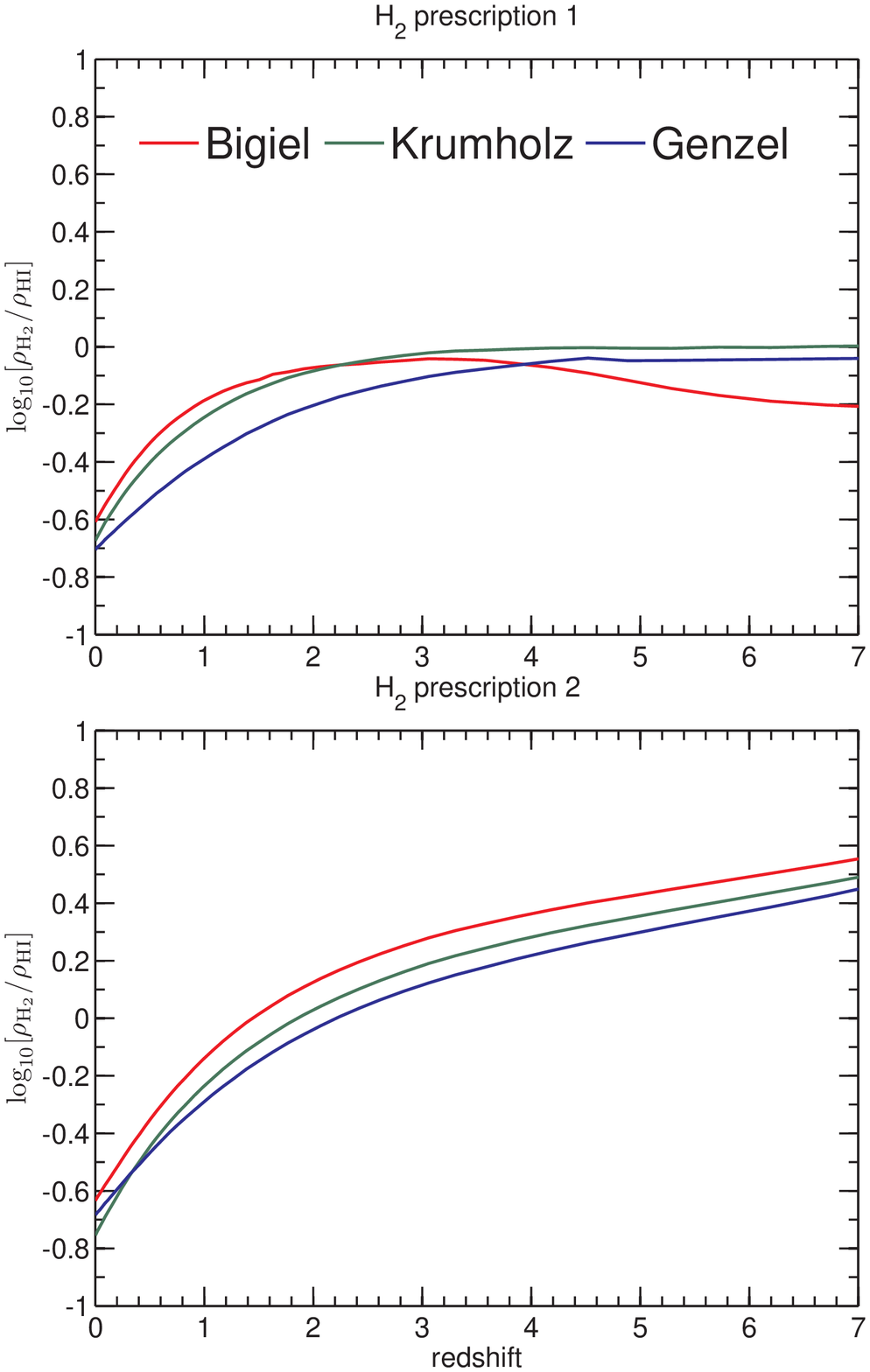}\\
 \caption{ Bottom right subfigure: The redshift evolution of the cosmic molecular-to-atomic gas ratio $\rho_{\h2}/\rho_{\rm HI}$ in the models. The top and bottom panels show results for the two $\h2$ fraction prescriptions. In each panel, three different coloured curves show for the star formation models, as labelled. Top and bottom left subfigures: The atomic gas to stellar mass ratio $\HIs$, the molecular gas to stellar mass ratio, and the molecular-to-atomic gas ratio $\Hs$, is plotted as a function of stellar mass $M_*$ for model galaxies at redshifts $z=0, 1, 2, 3, 4$. In each subfigure, the left column shows results for $\h2$ prescription 1 and the right column shows results for $\h2$ prescription 2. In each panel, the curves are the mean values from the galaxies in each stellar mass bin, and the colours represent results at different redshift, as indicated at the top of the figure.
 } \label{fig:H2HIstar}
\end{figure*}

In the bottom-right section of Fig. \ref{fig:H2HIstar} , we plot the redshift evolution of the cosmic molecular-to-atomic gas mass ratio $\rho_{\h2}/\rho_{\rm HI}$ for the three star formation laws that involve molecular gas (Bigiel, Krumholz and Genzel), as well as for our two $\h2$ fraction prescriptions. Note that the cosmic densities are obtained by integrating over galaxies with stellar masses larger than $10^9\ms$. We see that for a given $\h2$ fraction prescription, the Bigiel, Genzel and Krumholz yield very similar results. On the other hand, predictions for the two $\h2$ fraction prescriptions do diverge quite strongly at high redshifts. For $\h2$ prescription 1, $\rho_{\h2}/\rho_{\rm HI}$ reaches a maximum at $z\sim3$, and then either remains flat or decreases towards higher redshifts. For $\h2$ prescription 2, $\rho_{\h2}/\rho_{\rm HI}$ increase monotonically with redshift $z$.

Recall that for $\h2$ prescription 2, the molecular-to-atomic gas ratio is determined by interstellar pressure, which
depends on both gas surface density and stellar surface density (see Equation 31 and 32 in Fu10). At high redshifts, gas and stellar densities in disks are higher and this caused $\rho_{\h2}/\rho_{\rm HI}$ to increase strongly with redshift. For $\h2$ prescription 1, the molecular-to-atomic gas ratio is set by the combination of gas surface density and gas metallicity, because the Krumholz et al. (2009b) models show that $\h2$ molecules form very inefficiently in low metallicity gas. Because the gas-phase metallicity of galaxies decreases at higher redshifts, the formation of $\h2$ is suppressed and the molecular-to-atomic ratio evolves more weakly.

We note that the evolution of molecular-to-atomic ratio has been studied in previous work. Obreschkow \& Rawlings (2009) and Power et al. (2010) both adopted a pressure-based $\h2$ fraction prescription and obtained at monotonic increase in $\rho_{\h2}/\rho_{\rm HI}$ ratio as a function of redshift, in agreement with our results for $\h2$ fraction prescription 2. Lagos et al. (2011b) calculated molecular-to-atomic ratio for both $\h2$ prescriptions. They again predict a monotonic increase in $\h2$/HI ratio if they adopt a pressure-based $\h2$ prescription based on the model from Baugh et al. (2005). However, their preferred model adopts the pressure-based $\h2$ prescription with the model from Bower et al. (2006, Bow06.BR) predicts a slow {\em decrease} in $\h2$/HI ratio at $z>4$.
They state that this is caused by galaxies in haloes with masses $M_{\rm halo}<10^{11}\ms$) at $z>4$, where HI dominates. We note that
our predictions are for galaxies with stellar masses larger than $10^9\ms$, so we do not actually integrate over these very low mass halos. It is likely that observations of gas in high redshift galaxies in the near-term future will be restricted to fairly massive/luminous systems where molecules will still be able to form.

In order to estimate cosmic mass densities of HI or $\h2$, one either requires ``blind'' surveys to detect the HI and CO line over wide areas of the sky and over a significant range in frequency space, or one must target complete samples of galaxies spanning a wide range in stellar mass/optical luminosity. This will not be possible for some time. Observers are presently assembling restricted samples of high redshift galaxies with CO line measurements (e.g. Tacconi et al. 2010; Daddi et al. 2010; Riechers et al. 2010; Geach et al. 2011). Most of the samples have been selected using indicators sensitive to star formation rate, so do not represent ``fair'' samples of galaxies of given stellar mass. In the near term future, it will be possible to observe reasonable large stellar-mass selected samples of high redshift galaxies with ALMA.

In top and bottom-left subfigures of Fig. \ref{fig:H2HIstar}, we plot the molecular gas-to-stellar mass ratio $\Hs$, the atomic gas-to-stellar mass ratios $\HIs$, and the molecular-to-atomic gas mass ratio $\HHI$ as functions of stellar mass for galaxies at redshifts 0, 1, 2, 3 and 4. Results are shown for the three star formation models that involve molecular gas. In each subfigure, the top panels
show $\Hs$ vs $M_*$, the middle panels shows $\HIs$ vs $M_*$ and the bottom panels show $\HHI$ vs $M_*$. The two columns represent the two $\h2$ prescriptions. The curves in each panel represent the mean value for the galaxies in each stellar mass bin.

We see that both $\Hs$ and $\HIs$ decrease from high redshift to the present day. As the universe evolves, more stars are formed in galaxies and gas accretion rates decline. This leads to a decline in $\Hs$ and $\HIs$ with cosmic time. molecular gas fraction $\Hs$ always evolves more strongly than the atomic gas fraction $\HIs$. This is because galaxies are denser at higher redshifts, so molecular-to-atomic gas ratios are higher, as shown in the bottom panels. The strongest evolution in gas fraction occurs for the Bigiel star formation model. This is because the timescale for gas to be consumed into stars remains constant with redshift. The Genzel model produces the weakest redshift evolution, because of the $1/t_{\rm dyn}$ scaling.

The evolution in molecular-to-atomic fraction is predicted to be similar for galaxies of all stellar masses over the range $10^9$ to $10^{10.5} \ms$. Interestingly, the Bigiel star formation model with the Krumholz, Mckee \& Tumlinson (2009b) $\h2$-HI conversion prescription yields the {\em largest} increase in molecular-to-atomic ratio from $z=0$ to $z=1$, and the {\em smallest} increase from $z=3$ to $z=4$. $\HHI$ is predicted to increase by a factor of 5 from $z=0$ to $z=1$, but remain approximately constant from $z=3$ to $z=4$.
In contrast, the Genzel model predicts a more gradual and even evolution in $\HHI$ (a factor $\sim 2$ increase out to $z=1$).

In summary, the evolution of the molecular-to-atomic ratio does provide a way to distinguish between different $\h2$ prescriptions with future observational data. The fact that this evolution is independent of the stellar mass of the galaxy means that the test can be applied in the regime where the $X$ factor required to convert from CO luminosity to $\h2$ gas mass is not severely affected by metallicity effects (see next section).

\subsubsection{CO luminosity functions at high redshift: dependence on conversion factor}

Our models predict the molecular hydrogen content of galaxies, while observations only detect emission produced by CO molecules. To convert from CO luminosity to $\h2$ gas mass, a conversion factor $X$ must be assumed (Eq. \ref{eq:xco}). So far, we have assumed that this conversion factor is a constant $X=2\times10^{20}\rm{cm^{-2}~K^{-1}~km^{-1}~s}$ for all galaxies. Simulations of radiative transfer through molecular clouds have shown that $X$ will depend on the physical conditions in the ISM, including gas metallicity, temperature, velocity dispersion, and column density (e.g. Glover \& MacLow 2011). It has not been clear, however, how the predictions of ISM simulations on scales of a few parsecs translate to predictions of how the conversion factor averaged on much larger scales, or even over the whole galaxy, will depend on {\em global} galaxy physical parameters.

In recent work, Narayanan et al. (2011) and Feldmann et al. (2012) have coupled sub-grid models of the ISM, such as those produced by Glover \& MacLow (2011), to simulate the galaxy formation with cosmological initial conditions (Feldmann et al. 2012) or with simpler set-ups (Narayanan et al. 2011).

Feldmann et al. (2012) find that after smoothing on scales of $\sim1$ kpc, the dependence of the conversion factor on gas column density and interstellar UV flux is weak. They find that $X$ is mainly determined by the gas metallicity with the following power-law dependence
\begin{equation}\label{eq:xfeldmann}
\log_{10}X=a_1\log_{10}\left(Z/Z_{\odot}\right)+a_0
\end{equation}
in which $Z/Z_{\odot}$ is the gas metallicity relative to the solar value, and the unit of $X$ is $\rm{cm^{-2}~K^{-1}~km^{-1}~s}$.

In Narayanan et al. (2011), they test the effect of large-scale interstellar medium environment on the CO-$\h2$ conversion factor by coupling subgrid models of $\h2$ formation and a radiative transfer code to simulations of both quiescent galaxies and star burst galaxies formed in mergers. They find that $X$ in their model results \emph{on average} can be well fitted by the function:
\begin{equation}\label{eq:xnarayanan}
X=1.3\times10^{21}/(Z\Sigma_{\rm H_2}^{0.5})
\end{equation}
(Equation 10 of Narayanan et al. 2011). $X$ is in units of $\rm{cm^{-2}~K^{-1}~km^{-1}~s}$, $\Sh2$ is in units of $\mspc$,
and $Z$ is the gas metallicity relative to the solar value. $Z$ and $\Sh2$ are mass-weighted mean values for the whole galaxy disk.

In this section, we present model predictions for the evolution of the CO(1$\to$0) luminosity function from $z=0$ to high redshift.
Because the Bigiel star formation model has so far yielded results that are in best agreement with observations, we adopt Bigiel as our ``fiducial'' model in this section for simplicity.

Following Young \& Scoville (1991) and Obreschkow \& Rawlings (2009), we adopt a virial scaling for the CO line width to convert $\h2$ masses to CO luminosities. The CO emission $L_{\rm CO}$ is defined as $L_{\rm CO}\equiv4\pi D^2_{\rm 1}S_{\rm CO}$, in which $S_{\rm CO}$ is the CO($1\to0$) line flux and $D_{\rm l}$ is the luminosity distance. The relation between $\h2$ mass and CO flux is
\begin{multline}\label{eq:xlmh2}
\frac{M_{\h2}}{\ms}=\\
\frac{580X}{10^{20}\rm{cm^{-2}~K^{-1}~km^{-1}~s}}\left(\frac{\lambda}{\rm mm}\right)^2\frac{S_{\rm CO}}{\rm{Jy~km~s^{-1}}}\left(\frac{D_{\rm l}}{\rm Mpc}\right)^2
\end{multline}
in which $\lambda=2.6$mm is the wavelength of rest-frame CO($1\to0$) emission. So the relation between $L_{\rm CO}$ and $M_{\h2}$ is
\begin{multline}\label{eq:lco}
\frac{L_{\rm CO}}{\rm{Jy~km~s^{-1}~Mpc^2}}=\\
3.2\times10^{-3}\frac{M_{\h2}}{\ms}\left(\frac{X}{10^{20}\rm{cm^{-2}~K^{-1}~km^{-1}~s}}\right)^{-1}
\end{multline}

We present results for 3 different assumptions about the conversion factor $X$:
\begin {enumerate}
\item $X=2\times10^{20}\rm{cm^{-2}~K^{-1}~km^{-1}~s}$ for all galaxies
\item $\log_{10}X=a_1\log_{10}\left(Z/Z_{\odot}\right)+a_0$ from Feldmann et al. (2012). We adopt their results averaged on 4 kpc scale, i.e. $a_1=-0.66,~a_0=20.5$ in Eq. (\ref{eq:xfeldmann}). The conversion between $\h2$ and CO is calculated in each radial ring in the disk using the local gas metallicity and $\h2$ surface density. We then sum $L_{\rm CO}$ in each ring to get the total CO luminosity for the whole galaxy.
\item Eq. (\ref{eq:xnarayanan}): $X=1.3\times10^{21}/(Z\Sigma_{\rm H_2}^{0.5})$ from Narayanan et al. (2011).
$Z$ and $\Sigma_{\rm H_2}$ are mass-weighted mean values for the whole disk obtained by averaging
over all the rings.
\end {enumerate}

\begin{figure*}
\centering
 \includegraphics[angle=-90,scale=0.5]{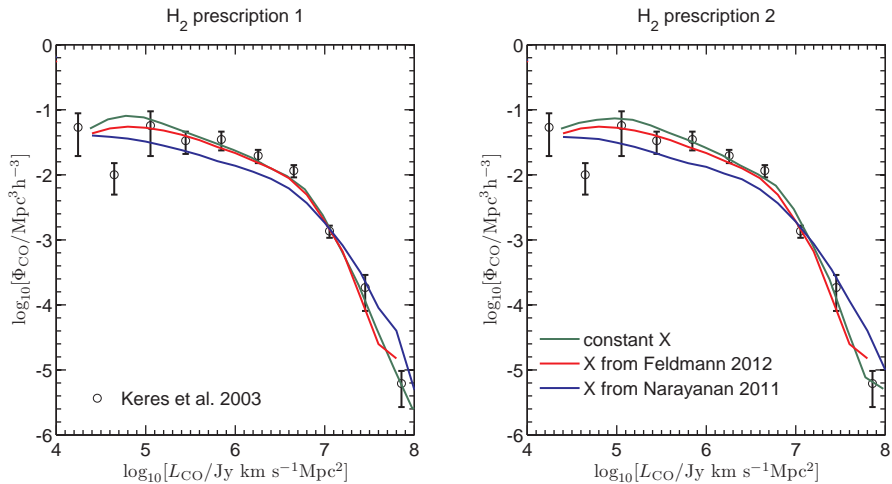}\\
 \caption{The CO luminosity functions are plotted at $z=0$ for the Bigiel star formation model. The two panels show results for the two
 $\h2$ fraction prescriptions. The green curves show results for a constant $X$ factor, the red curves show results for the $X$ factor prescription of Feldmann et al. (2012) and the blue curves show results for the $X$ factor prescription of Narayanan et al. (2011). Note that the model mass-metallicity relations have been ``corrected'' to have the same slope as the observations (see text). The black circles with error bars are observational results from Keres et al. (2003).
 } \label{fig:lcoz0}
\end{figure*}

In Fig. \ref{fig:lcoz0}, we show CO luminosity functions at $z=0$ for the Bigiel model and both $\h2$ fraction prescriptions. As discussed previously, the model MZ relation is not as steep as in the observations of Maiolino et al. (2008). To account for this, we have applied a correction to the gas metallicities in the model galaxies to bring the MZ relation into line with the data. We compute the CO luminosity function using these corrected metallicities. The red solid curves show results when the metallicity-dependent conversion factor in
Eq. (\ref{eq:xfeldmann}) is used to derive the CO luminosity, the blue curves are for conversion factor in Eq. (\ref{eq:xnarayanan}), and the green solid curves show results for the constant conversion factor. The observational results are from FCRAO as given in Keres et al. (2003).

Fig. \ref{fig:lcoz0} shows that the different $X$ factor prescriptions make only small difference to the CO luminosity function at $z=0$. The biggest differences are at the faint end, where the observational errors are large. The Narayanan et al's prescription does predict a larger number of very CO-luminous galaxies, but these are very rare and we caution that the error bars on the observational data do not include cosmic variance, which will dominate the error budget in this regime. We conclude that the current $z=0$ determination of the CO luminosity function is not able to discriminate between the different $X$ factor possibilities.

\begin{figure}
\centering
 \includegraphics[angle=0,scale=0.45]{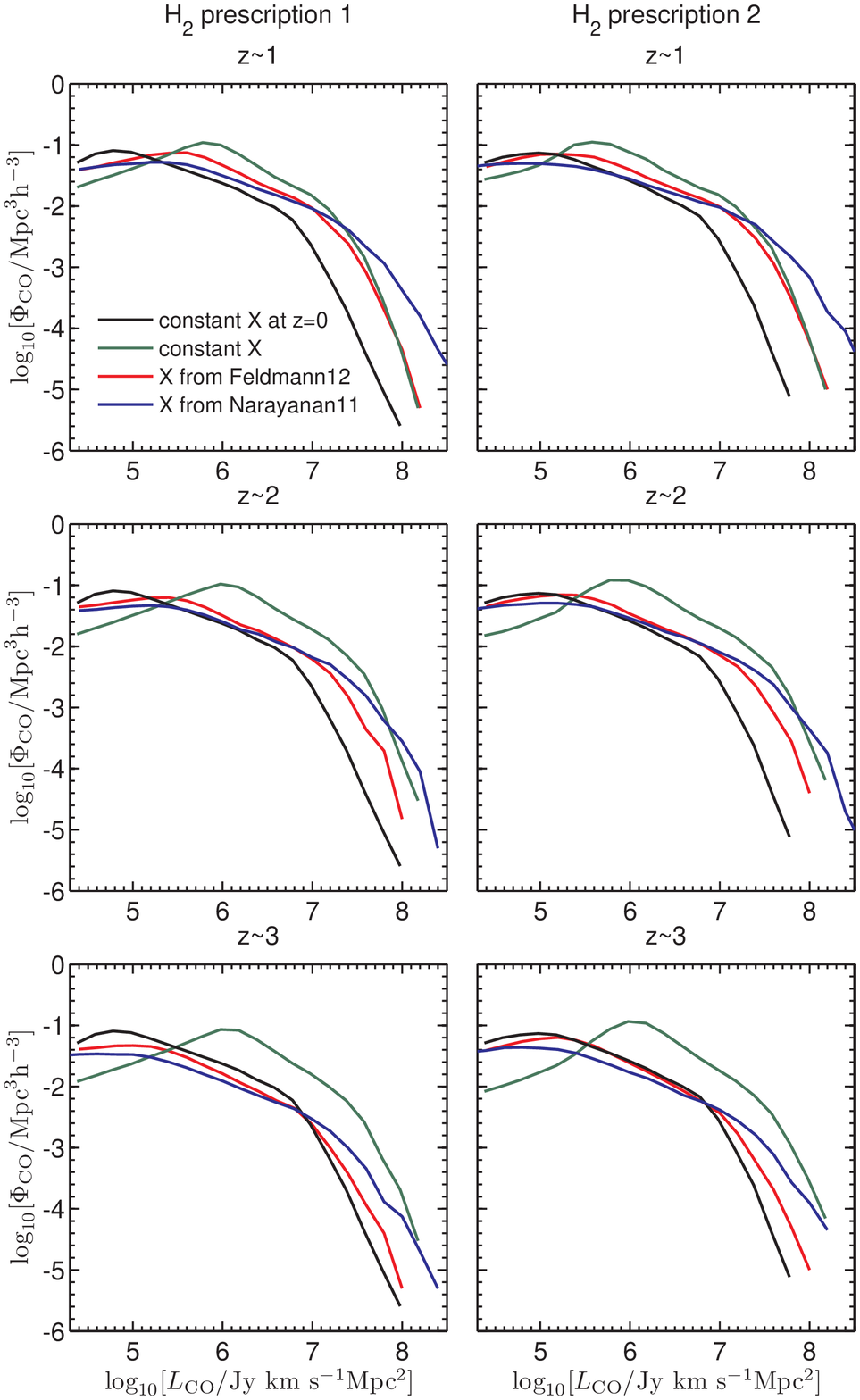}\\
 \caption{The CO luminosity functions at high redshift. The three rows show results at $z\sim1,2,3$ respectively and the two columns show results for the two $\h2$ fraction prescriptions. In each panel, the red, green and blue curves have the same meaning as in Fig. \ref{fig:lcoz0}. To guide the eye, we reproduce the $z=0$ CO luminosity function with constant $X$ in each panel (black curves).
 } \label{fig:lcoz}
\end{figure}

Fig. \ref{fig:lcoz} shows predicted CO luminosity functions at $z \sim1,~2$ and $3$ like in Fig. \ref{fig:lcoz0}. Once again, we have corrected the model MZ relations to match the observed ones. Red, blue and green curves are for the Feldmann, Narayanan and constant conversion factors. To guide the eye, we also plot the $z=0$ CO model luminosity function with constant $X$ factor in black. Fig. \ref{fig:lcoz} shows that the high-z predictions are much more sensitive to the X-factor prescription. If we apply a constant conversion factor, the bright end of the CO luminosity function evolves strongly out to $z \sim 2$ and then remains constant to $z \sim 3$. If we apply the metallicity-dependent conversion factor of Feldmann et al. (2012), the bright end evolves very little from $z=1$ to $z=2$, and then {\em decreases} to $z=3$. The Narayanan et al. (2011) prescription yields evolution that is intermediate between the two.

\subsection{The redshift evolution of the stellar mass function} \label{chap:smfz}

We have shown that the Bigiel star formation model leads to a cosmic star formation history that is peaked at somewhat lower redshifts than that obtained with the Kennicutt model. In addition, the Bigiel model produces a mass-metallicity relation that evolves more strongly as a function of redshift and is in better agreement with observations.

In this section, we investigate the effect of the Bigiel star formation model on the redshift evolution of the stellar mass function. Guo11 show that their model correctly fits the abundance of the most massive galaxies out to $z\sim$4, but over predicts the abundance of galaxies with stellar masses less than $10^{10}\ms$ at redshifts greater than 0.6. Lower mass galaxies clearly complete their formation
too early in the model (see also Fontanot et al. 2009; Weinmann et al. 2011).

\begin{figure*}
\centering
 \includegraphics[angle=-90,scale=0.65]{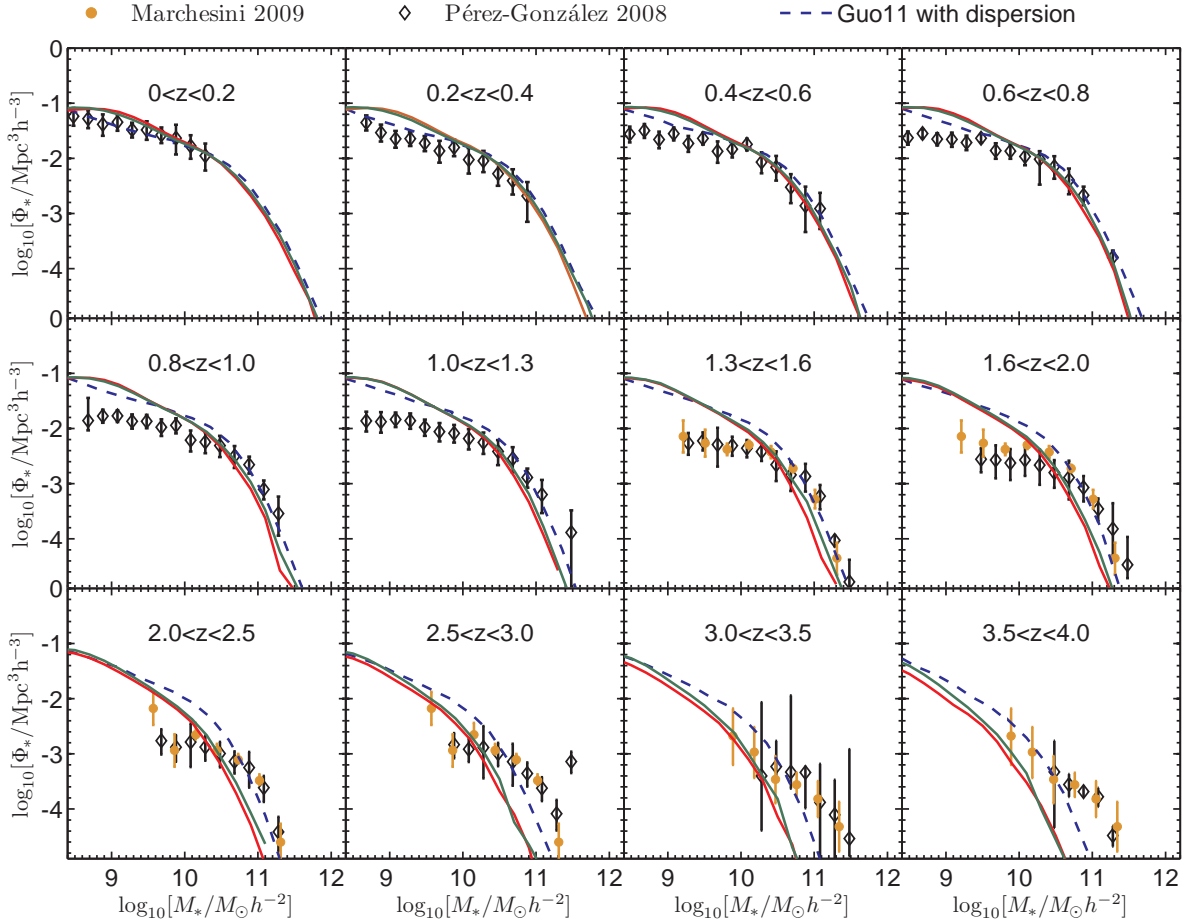}\\
 \caption{ Model stellar mass functions in different redshift bins. Each panel represents a redshift bin. The observational results of P{\'e}rez-Gonz{\'a}lez et al. (2008) are shown as black diamonds with error bars. The orange dots with error bars are from Marchesini et al. (2009). The redshift bins in Marchesini et al. (2009) are wider than in P{\'e}rez-Gonz{\'a}lez et al. (2008), so two consecutive panel may contain the same orange dots. The solid red and green show Bigiel model stellar mass functions for $\h2$ fraction prescription 1 and 2. The blue dashed curves are results from Guo11. All the model curves have been convolved with a gaussian of dispersion 0.25 dex in $\log M_*$ to simulate the effect of errors in the observed stellar mass estimates.
 } \label{fig:smfz1}
\end{figure*}

In Fig. \ref{fig:smfz1}, we plot stellar mass functions in different redshift bins for the Bigiel star formation model. The observational data is taken from P{\'e}rez-Gonz{\'a}lez et al. (2008) and Marchesini et al. (2009). Stellar masses have been derived assuming a Chabrier IMF (Chabrier 2003). The red and green solid curves show model results for the two $\h2$ fraction prescriptions. The model stellar mass functions are convolved with a Gaussian of dispersion 0.25 dex in $\log M_*$ to account for the errors in the observed stellar mass estimates. We also plot the model results from Guo11 for comparison.

Fig. \ref{fig:smfz1} makes it clear that the Bigiel star formation law does not solve the mismatch with the observed redshift evolution of the stellar mass function. In fact, the match to the massive end of the mass function at high redshifts is worse than it was previously, and the problems at the low mass end persist. \footnote{We note that we have analyzed the the redshift evolution for the stellar mass functions for the other 3 star formation models in this paper. At $z\lesssim1.5$, the results are nearly identical. At higher redshifts, the KMT and Bigiel models, in which molecular gas is always converted into stars over a fixed timescale, yield very similar results. The results from the Kennicutt and Genzel models are similar to the results from Guo11 (the blue dashed curves in Fig. \ref{fig:smfz1}). In all three of these models, the cold gas consumption timescale $t_{\rm dyn}$ decreases at higher redshifts.}

To solve the problem of stellar mass function at both low and high mass ends will likely require changes to other aspects of the galaxy formation physics that are not considered in this paper, for example the model for how supernovae eject gas from galaxies and dark matter halos and how this gas is later re-incorporated back into the galaxy (Henriques, private communication).

\section{Summary and Discussion}

In this paper, we study the effect of different star formation models on the redshift evolution of interstellar metals, atomic and molecular gas in galaxies using semi-analytic models of galaxy formation embedded in the Millennium I and II simulations. In order to track the formation of molecular gas in galaxies, we adopt the methodology developed by Fu et al. (2010) in which each galactic disk is represented by a set of concentric rings. Two simple prescriptions for molecular gas formation are included in the models: one is based on the analytic calculations by Krumholz, McKee \& Tumlinson, and the other is a prescription where the $\h2$ fraction is determined by the pressure of the interstellar medium.

We analyze four different star formation models: (i) A model based on the results of Bigiel et al. (2008) where $\Sigma_{\rm SFR}$ scales with $\Sigma_{\h2}$ in regions of the disk where molecular gas dominates, and with $\sgas^2$ in regions of the disk where atomic gas dominates, (ii) A model based on the results of Kennicutt (1989, 1998) where $\Sigma_{\rm SFR}$ scales with $\sgas$ multiplied by the inverse of the disk dynamical time, (iii) A model motivated by a recent paper by Genzel et al. (2010) where $\Sigma_{\rm SFR}$ scales with $\Sh2$ multiplied by the inverse of the disk dynamical time, (iv) A model proposed by Krumholz et al. (2009b) where $\Sigma_{\rm SFR}$ scales with $\Sh2$ with a proportionality factor that depends on the surface density of molecular gas in the galaxy.

For every star formation model, we re-tune the free parameters of the semi-analytic model so that we fit the present-day stellar mass function, as well as the mass functions of both atomic and molecular gas at $z=0$. We then compare the redshift evolution of the cosmic star formation density, the cosmic $\h2$ to HI fraction, the relation between gas-phase metallicity and stellar mass, and the atomic and molecular gas mass fractions of galaxies for the four star formation models listed above.

Our two main conclusions are the following:

\noindent (i) The two $\h2$ fraction prescriptions yield very different predictions for the cosmic value of $\rho_{\h2}/\rho_{\rm HI}$ at $z>3$. In Fu10, it was shown that two $\h2$ prescriptions predict almost identical results at $z=0$ for galaxies with $M_*>10^9\ms$. In this paper, we show that $\rho_{\h2}/\rho_{\rm HI}$ increases monotonically with redshift for the pressure-based prescription, but decreases at $z>3$ for the KMT prescription because higher redshift galaxies have lower gas-phase metallicities.

\noindent (ii) The timescale over which gas is converted into stars has a large effect on the redshift evolution of the stellar mass-gas metallicity relation. The $1/t_{\rm dyn}$ dependence in the Kennicutt and Genzel star formation models leads to a decrease in gas consumption time with redshift proportional to $(1+z)^{-3/2}$. This produces mass-metallicity relations that do not evolve with redshift,
in contradiction with observations. The Bigiel star formation model in which star formation depends only on the surface density of molecular gas yields results in better agreement with observations.

\subsection {Relation to previous work}

We will now discuss the results in our paper in the context of other recent theoretical analyses of the evolution of gas and metals in galaxies. We have shown that the predicted evolution is very sensitive to assumed amount of infall of unenriched gas from the external environment, so we will confine our discussion to models where the evolution of the infall rates is in accord with expectations from the $\Lambda$CDM "concordance" cosmological model. Other models exist in the literature where infall is parameterized by free functional forms (e.g. Calura et al. 2009).

Sakstein et al. (2011) study the evolution of the mass-metallicity relation using the GALICs semi-analytic models. They vary the star formation model while holding other parameters fixed. They also conclude that the evolution of the mass-metallicity relation is very sensitive to the assumed star formation mode, but find a preferred star formation models of the form $\rm{SFR}= \alpha \left( M_{\rm cold}/t_{\rm dyn}\right)(1+z)^{\beta}$, with $\beta=1$. In this model, therefore, the conversion time between gas and stars is even
{\em shorter} at high redshifts than in our Kennicutt model. The conclusions of this paper thus appear to be in strong disagreement with
ours. Other physical inputs such as the prescription for supernova feedback appear to be very similar to ours, so the discrepancy is puzzling. One clear difference in approach is that Sakstein et al. vary the free parameters of the models and study the effect {\em only} on the mass-metallicity relation. There is no discussion about whether their preferred model is able to fit other observational data, such as stellar and gas mass fractions.

Lagos et al. (2011a, 2011b) study the evolution of both atomic and molecular cold gas in galaxies as well as the relation between star formation rate and stellar mass using semi-analytic models. They again vary the star formation model, while holding other parameters fixed. Using the distribution of galaxies on the active and passive branches of the SFR vs $M_*$ diagram at $z=0$ , the HI mass function at $z=0$, and the K-band luminosity functions at $z>0.5$, Lagos et al. arrive at a ``preferred model'' with the star formation model from Leroy et al. (2008) and $\h2$ prescription from Blitz and Rosolowsky (2006), which is very similar to our preferred ``Bigiel'' model,
but their reasons are quite different. The evolution of the mass-metallicity relation is not considered in their papers. Lagos et al. claim that the SFR vs $M_*$ plot places the strongest constraint on the star formation models. We would argue that the AGN feedback prescription
is most important in determining the relative fraction of galaxies on the active and passive branches of this relation (Kauffmann et al. 2012). We also note that Lagos et al. choose not to re-normalize their model parameters to fit $z=0$ stellar and gas mass fractions when they change their star formation model, but retain the values in Baugh et al. (2005) and Bower et al. (2006). This difference in approach makes it difficult to compare our results with those of Lagos et al. in a quantitative way.

The evolution of gas and metals in galaxies has also been studied using cosmological hydrodynamical simulations. Dav\'e, Finlator \& Oppenheimer (2011) study the evolution of the mass-metallicity and the cold gas mass versus stellar mass relations to high redshift using cosmological hydrodynamical simulations. They run a suite of models in which the star formation model is held fixed, but the model for winds driven by supernova explosions and radiation input from young stars varies. They find that the supernova feedback mainly affects the {\em shapes} of these two relations. The {\em average evolution} in the two relations with redshift is similar for all their models. The authors note, however, that the degree to which the metallicities of galaxies of different masses will evolve is sensitive to the wind model. For example in the model with no winds, the metallicities of low mass galaxies evolve more strongly than those of high mass galaxies; the opposite is true for the ``momentum-driven'' feedback model. In a recent paper, Duffy et al. (2012) predict the evolution of $\Hs$ and $\HIs$ using the OverWhelmingly Large Simulations (OWLS) project. At redshift zero, their simulations appear to fit observed trends in HI gas mass fraction as a function of stellar mass quite well, but the predicted molecular gas fractions do not agree with data.
The simulations that include AGN feedback produce molecular gas fractions that are too low by a factor of more than 100 (see their Figure 6).

In this paper, we vary only the star formation model, leaving the supernova feedback model fixed. This changes the {\em timescale} over which gas that is accreted onto a galaxy is converted into stars, but does not affect the {\em quantity of gas} that is accreted as function of dark matter halo mass or of redshift. Because gas accretion rates are predicted to be large at high redshifts, the average gas-phase metallicity of high-z galaxies is very sensitive to the adopted gas conversion timescale. We have concluded that the Bigiel star formation model best reproduces the average evolution of the mass-metallicity relation, but we caution that our models do not yet fit the detailed shape of the MZ relation, nor are they able to fit the observed redshift evolution of the stellar mass function. In future work,
we plan to vary star formation, gas ejection and gas re-accretion prescriptions concurrently and compare with a wider range of observational diagnostics of the high redshift population of star-forming galaxies.

We have also not attempted any quantitative comparison between the gas properties of high redshift galaxies and the predictions of our models, believing this to be premature. The available samples are too small and the selection effects too uncertain for robust comparisons at this point, but we expect the situation to change soon.

\def\apj{ApJ}
\def\apjl{ApJL}
\def\apjs{ApJS}
\def\aj{AJ}
\def\aap{A\&A}
\def\araa{ARA\&A}
\def\aapss{A\&AS}
\def\mnras{MNRAS}
\def\nature{Nature}
\def\apss{Ap\&SS}
\def\pasp{PASP}

{} 

\end{document}